\documentclass[preprint,aps,superscriptaddress,nofootinbib,longbibliography]{revtex4-2} % For arXiv

\usepackage{type1cm}
\usepackage{graphicx}
\graphicspath{{./Figures/}}

\usepackage{dcolumn}
\usepackage{bm}
\usepackage{nicefrac}
\usepackage{upgreek} % For adiabat use upalpha. For alpha-particle, use regular alpha
\usepackage{xcolor}
\usepackage{bigints}

\usepackage[utf8]{inputenc}
\usepackage[T1]{fontenc}
\usepackage{mathptmx}

\usepackage{etoolbox}
\usepackage[export]{adjustbox}

\usepackage{orcidlink}

\usepackage{hyperref}
\hypersetup{hidelinks}

\begin{document}

\frenchspacing
\title{A Hybrid Scheme to Achieve Highest Implosion Performance on the OMEGA Laser}
\author{P. S. Farmakis\orcidlink{0000-0002-6520-355X}}
\email{pfar@lle.rochester.edu}
\affiliation{Laboratory for Laser Energetics, University of Rochester, Rochester, New York 14623-1299, USA}
\affiliation{Department of Mechanical Engineering, University of Rochester, Rochester, New York 14627, USA}
\author{L. Ceurvorst\orcidlink{0000-0002-4215-3326}}
\affiliation{Laboratory for Laser Energetics, University of Rochester, Rochester, New York 14623-1299, USA}
\author{V. Gopalaswamy\orcidlink{0000-0002-8013-9314}}
\affiliation{Laboratory for Laser Energetics, University of Rochester, Rochester, New York 14623-1299, USA}
\affiliation{Department of Mechanical Engineering, University of Rochester, Rochester, New York 14627, USA}
\author{D. Cao\orcidlink{0000-0003-1085-2322}}
\affiliation{Laboratory for Laser Energetics, University of Rochester, Rochester, New York 14623-1299, USA}
\author{K. S. Anderson\orcidlink{0000-0002-9248-4545}}
\affiliation{Laboratory for Laser Energetics, University of Rochester, Rochester, New York 14623-1299, USA}
\author{A. Lees\orcidlink{0000-0003-4573-4035}}
\affiliation{Laboratory for Laser Energetics, University of Rochester, Rochester, New York 14623-1299, USA}
\affiliation{Department of Mechanical Engineering, University of Rochester, Rochester, New York 14627, USA}
\author{D. Patel\orcidlink{0000-0003-1204-4468}}
\affiliation{Laboratory for Laser Energetics, University of Rochester, Rochester, New York 14623-1299, USA}
\affiliation{Department of Mechanical Engineering, University of Rochester, Rochester, New York 14627, USA}
\author{R. Betti\orcidlink{0000-0001-8742-0304}}
\affiliation{Laboratory for Laser Energetics, University of Rochester, Rochester, New York 14623-1299, USA}
\affiliation{Department of Mechanical Engineering, University of Rochester, Rochester, New York 14627, USA}
\affiliation{Department of Physics and Astronomy, University of Rochester, Rochester, NY USA}
\date{\today}
\begin{abstract}
\noindent Merging direct and indirect-drive has long been viewed as an optimal hybrid laser-fusion scheme that combines the uniformity of x rays with the efficiency of direct illumination. We present the first integrated 2D simulations of hybrid shock drive (HSD) targets for the OMEGA laser. The HSD scheme [L. Ceurvorst \textit{et al.}, Phys. Rev. E \textbf{101} 063207 (2020)] uses x rays from a thin Au-coated x-ray converter outer shell to drive the initial shock into a standard direct-drive capsule. Direct illumination is used to implode the target after the first shock. The design effectively suppresses laser-imprint seeding of hydrodynamic instabilities, maintaining shell integrity during the implosion. This scheme will enable fielding low-adiabat, high-convergence implosions on OMEGA with expected performance greatly exceeding those of current designs. HSD targets are projected to significantly enhance fusion yields, potentially increasing the record Lawson parameter by $\sim$85\% on OMEGA while effectively eliminating the requirement for laser smoothing. These results position HSD as a robust platform for high-performance implosions, paving the way for advanced high-gain inertial fusion energy targets.
\end{abstract}
\maketitle

Laser direct drive (LDD) and laser indirect drive (LID) \cite{nuckolls1972laser,craxton2015direct} are the two major laser-fusion approaches to inertial confinement fusion (ICF). In both schemes, a shell containing deuterium--tritium (DT) fuel is imploded by ablation pressure, produced either by direct laser illumination in LDD or by x rays in LID. These approaches represent complementary limits: LDD favors efficient coupling, while LID favors drive uniformity.

In LID, laser beams grouped into rings at fixed polar angles irradiate distinct regions of the inner wall of a high-$Z$ radiation cavity, or hohlraum. The wall converts laser energy into x rays that ablate the capsule surface. This conversion reduces the energy available to the implosion. A typical ignition-scale NIF capsule absorbs $\sim$200--250~kJ from a 2-MJ laser drive \cite{zylstra2021metrics}. The advantage is a radiation field largely free of small-scale nonuniformities, although low-mode asymmetries remain and require tuning of beam power balance and inter-cone wavelength shifts \cite{kritcher2018energy,marozas2018wavelength}. LDD, by contrast, couples energy more efficiently to the capsule. OMEGA’s 30-kJ LDD implosions reach $\sim$70\% coupling efficiency \cite{williams2024demonstration}, with projections exceeding $90\%$ for broadband lasers \cite{campbell2020direct}. Direct illumination also provides nearly spherical beam geometry, avoiding the intrinsic low-mode asymmetries associated with indirect drive. Beam-geometry ``mid-modes\rlap{,}'' $6<\ell<30$, can still arise but are often controlled through beam positioning. The dominant limitation is instead laser imprint, which seeds small-scale nonuniformities from individual-beam speckle patterns \cite{bodner1974rayleigh,ishizaki1997propagation,ishizaki1998model,bourgeade2012time}.

Therefore, the optimal drive of an ICF implosion would be a ``hybrid'' of direct and indirect-drive, combining the benefits of LDD (high coupling efficiency and good low-mode symmetry) and LID (free of high modes from laser imprint). Although this conclusion is straightforward, no integrated design under realistic conditions has yet demonstrated that a hybrid target can both generate a sufficiently intense, symmetric x-ray source to create an imprint-mitigating plasma atmosphere and allow later-time direct laser propagation to the capsule \cite{bocher1984improvement,emery1991hydrodynamic,eliezer1992,afsharrad1994,desselberger1995,dunne1995}. The concept developed at the Laboratory for Laser Energetics (LLE) uses x rays generated by direct illumination of a Au-coated thin plastic membrane to launch a uniform first shock in a DT-layered capsule. The ablated plasma formed by this shock then shields the capsule from laser imprint. This hybrid shock drive (HSD) scheme \cite{farmakis2023foam} was validated in planar experiments on OMEGA from 2016 to 2020 \cite{ceurvorst2020hybrid}.

An HSD variant was recently proposed by Thomas et al. \cite{thomas2024hybrid}, using circular baffles inside a cylindrical hohlraum to shape a uniform x-ray drive in a two-sided illumination geometry. That study, however, relied on a prescribed x-ray source from view-factor estimates rather than self-consistent hohlraum modeling, ignoring wall motion and LEH-window interactions. It also neglected laser imprint, assuming that thermal smoothing would suppress ablation-pressure nonuniformities.

Another approach by the Naval Research Laboratory (NRL) \cite{karasik2015suppression,karasik2021order} utilizes a high-$Z$ overcoat as a weak x-ray source to gently preheat and ablate a direct-drive target over many nanoseconds. The aim is to produce a plasma buffer between the laser deposition and the ablation front to smooth out laser imprint. NRL's target design has been limited to a simplified 1D regime \cite{obenschain2002effects,mostovych2008enchanced}, and the concept remains unrealistic for current laser facilities primarily due to the inability to generate such pulses.

\begin{figure}[tbp]
  \centering
  \begin{minipage}{0.99\columnwidth}
    \centering
    \includegraphics[width=\textwidth]{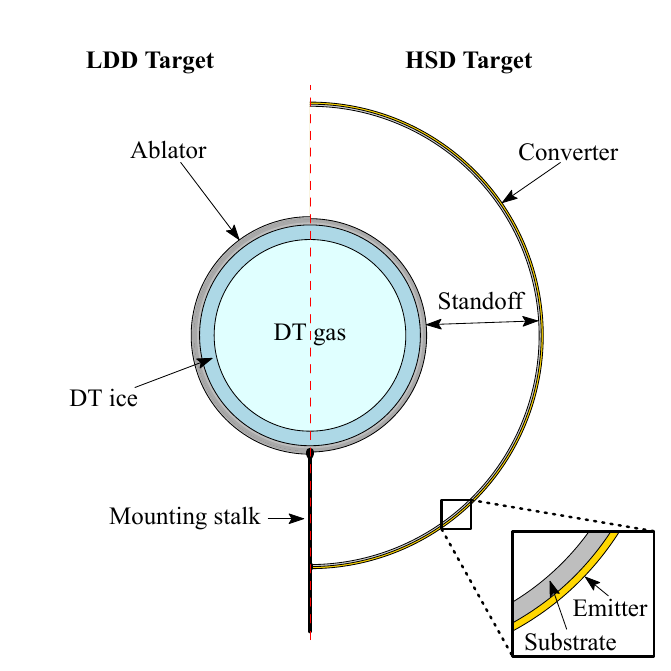}
    \caption{Target sketch showing a side-by-side comparison of the standard LDD ``bare'' target based on the OMEGA best-performers (left) and its mass-equivalent HSD (right). The layers (not to scale) consist of the converter (high-\textit{Z} emitter supported by a substrate, see inset) and offset from the cryogenic capsule (ablator and fuel).} 
    \label{fig:LDD_HSD_targets}
  \end{minipage}
\end{figure}

In this Letter, we present the first integrated simulations of hybrid-drive targets for high-performance implosions on OMEGA (see Fig.~\ref{fig:LDD_HSD_targets}). The simulations self-consistently model x-ray generation, the radiation-driven shock, and subsequent direct-drive acceleration. The HSD target consists of a thin Au-coated plastic converter shell surrounding a standard OMEGA DT-layered capsule, separated by either vacuum or a low-density gas buffer. We find that HSD substantially outperforms standard laser direct drive (LDD) by mitigating laser imprint and enabling low-adiabat, high-convergence implosions. On OMEGA, HSD achieves a near-twofold increase ($\sim$85\%) in the Lawson parameter, which translates to at least a fourfold reduction in the laser energy required for ignition. The high-$Z$ coating also provides intrinsic thermal shielding for capsule insertion into inertial-fusion-energy (IFE) reactor chambers \cite{blink1985high,moir1994hylife,sethian2010science}.

To assess the performance improvement from hybrid drive on an equal footing, we compare low-adiabat $(\alpha\sim2)$ HSD targets with a corresponding standard ``bare'' LDD target driven to the same adiabat and implosion velocity. The target and pulse designs are developed with the 1D radiation-hydrodynamic code \textsc{lilac} \cite{delettrez1976report}, and their multidimensional performance is evaluated with 2D \textsc{draco} simulations \cite{radha2005multi,marozas2006polar,marozas2018wavelength}. These calculations also determine the illumination requirements needed to produce a sufficiently uniform x-ray drive in a realistic OMEGA geometry.

\textit{Design of the hybrid target}--- The primary target components are illustrated in Fig. \ref{fig:LDD_HSD_targets}. The HSD design (right half) consists of a DT-layered capsule enclosed by a converter shell. This target is a mass-equivalent derivative of high-performance LLE designs. It features an outer CHSi-doped ablator to enhance laser absorption and mitigate laser-plasma instabilities \cite{seka2009two,solodov2022hot} during the main drive, and an inner CD layer to mitigate fuel preheat from silicon self-emission and mix \cite{follet2016two,gopalaswamy2022increasing}. Unlike current OMEGA implosions, the HSD target is driven on a low adiabat to achieve high convergence, high pressures, and high areal densities. This specific low-adiabat hybrid design configuration belongs to the same family of implosions as contemporary high-performance, high-adiabat OMEGA implosions \cite{nora2014theory} which enables a direct performance comparison for ignition scaling. These high-adiabat designs have achieved record Lawson parameters of $\chi^{\rm{no}\mathrel{-}\upalpha}_{\Omega}\simeq 0.2$ with neutron yields of $\simeq 2\times 10^{14}$, areal densities of $\simeq 160$ mg/cm$^2$ and stagnating DT mass of $\simeq 10$ $\mu g$ \cite{lees2021experimentally}. When hydrodynamically scaled to 2 MJ of laser energy, such implosions reach the burning plasma regime with energy gains close to unity \cite{gopalaswamy2024demonstration,williams2024demonstration,ceurvorst2025progress,less2025approaching}.

Laser systems like OMEGA and the NIF operate at 351-nm wavelengths allowing laser light to couple deep within the coronal plasma. This provides good hydrodynamic efficiency \cite{bodner1981critical} but leads to a narrow conduction zone separating the critical and ablation surfaces \cite{gardner1981wavelength}. At early times, when the conduction zone is very thin, thermal smoothing cannot mitigate nonuniformities in the ablation pressure seeded by the laser speckles \cite{betti1998growth}, resulting in a highly perturbed first shock \cite{goncharov2000model}. High-performance targets opted for high adiabats because they reduce the growth rates of these perturbations \cite{goncharov2014improving}, but limit overall performance \cite{craxton2015direct,goncharov2008modeling}. This physical picture shows why directly driven targets would benefit from driving the first shock with a smooth radiation flash, particularly at low adiabats.
\begin{figure}[!htbp]
  \centering
  \begin{minipage}{0.95\columnwidth}
    \centering
    \includegraphics[width=\textwidth]{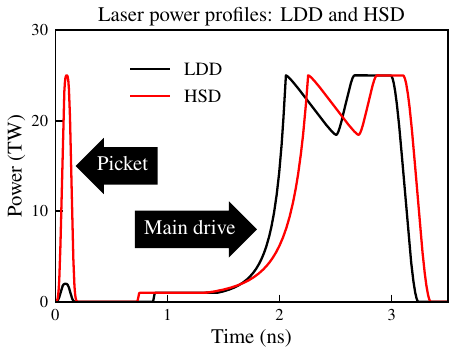}
    \caption{Laser pulse shapes for $\alpha\simeq 2$ used in the HSD target design (red) and standard LDD (black).}
    \label{fig:pulse_shape}
  \end{minipage}
\end{figure}
\begin{figure}[!htbp]
  \centering
  \begin{minipage}{\columnwidth}
    \centering
    \includegraphics[width=\linewidth]{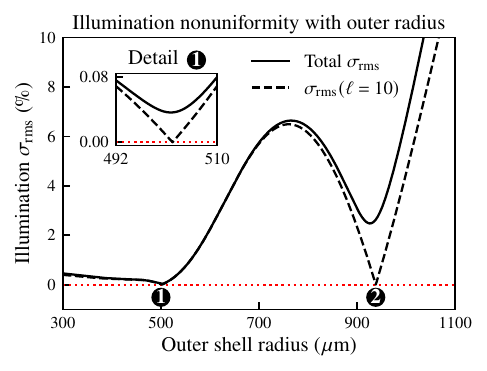}
    \caption{Illumination uniformity for a range of radii ($300$ to $1100\;\mu$m) using the hard sphere projection from \cite{skupsky1983uniformity}. It assumes the standard OMEGA (SG5-850) phase plates ($R_{95\%}\simeq425\;\mu$m, super-Gaussian exponent $\eta\simeq4.4$) at $t=0$. The markers (1 and 2) indicate radii with minimized nonuniformities; (1) is the cryogenic capsule, and (2) is the converter. The inset details the nonuniformity profile near the capsule.}
    \label{fig:skupsky_modes}
  \end{minipage}
\end{figure}

A smooth first shock can be accomplished using x rays uniformly emitted by a carefully designed source. For spherical implosions, this can be a spherical, micron-thick  converter shell with a high-$Z$ coating located several hundred microns from the ablator of the nested capsule. The converter absorbs the laser light and is rapidly heated up to $\sim$100~eV, emitting x rays. The x rays absorbed by the capsule generate Mbar-range ablation pressure on the capsule surface, driving the first shock. The standoff distance controls the geometrical smoothing effect of the x-ray drive on the capsule's ablator \cite{murakami1986smoothing,caruso1991quality,pollaine2000quality}.

Placing the converter layer between the laser and capsule creates a radiative cavity with several design consequences. The first is energetics. The converter is a tens-of-nanometers-thick Au-coating illuminated by the initial ``strong'' picket of the laser pulse (Fig.~\ref{fig:pulse_shape}). As the converter disassembles, roughly 3\% of the picket energy is deposited on the capsule as x rays. Generating Mbar x-ray pressures requires a picket energy of $\sim$1.5~kJ. Reducing the converter--capsule standoff improves radiative coupling, but increases the risk of \textit{hydrodynamic coupling}, in which laser-imprinted modes on the converter are transferred by nonuniform converter flows to the capsule surface, negating the benefit of a smooth x-ray shock. Larger standoffs hydrodynamically decouple the converter from the capsule, but increase the cavity volume and weaken x-ray coupling, limiting control of the first-shock strength and fuel adiabat \cite{ceurvorst2020hybrid}. In an energy-limited facility such as OMEGA \cite{soures1996direct}, this tradeoff can become impractical. Symmetry control is further complicated by finite-beam effects.

Another concern is mechanical support. Making a high-\textit{Z} converter shell self-supporting at the required radii would add considerable mass to the target. Our simulations in 1D and 2D show this increases the risk of x-ray preheat of the fuel later in the implosion, and complicates direct illumination of the capsule. Therefore, the amount of high-$Z$ material in the converter must be kept to a minimum. Structural integrity is provided by the converter's substrate, a thin CH shell. Its thickness is tuned iteratively to mitigate laser burnthrough at the falling edge of the picket pulse, when most of the converter material has been ablated off. The substrate shields the target by absorbing the remainder of the unconverted laser pulse so that laser imprint from the picket does not reach the capsule before a thick plasma atmosphere is in place. Simulations in \textsc{draco} have evaluated the effect of an unconverted laser picket tail on the capsule with an optimized substrate. They show negligible reduction in yield and areal density; less than $2\%$ and $3\%$, respectively.

A downside of thin high-$Z$ converters is that they provide high peak brightness and a hard x-ray spectrum in a quick flash, but rapidly lose internal energy to negative $PdV$ work below the radiative threshold. For solid Au, the critical radiation temperature $T_{\rm cr}\simeq110$ eV separates the radiative and hydrodynamic regimes \cite{atzeni2004physics}; our simulations give $T_{\rm max}\sim80$ eV at shock breakout which entails hydrodynamic expansion.

Previous work \cite{murakami1986smoothing,caruso1991quality,pollaine2000quality} has shown that controlling x-ray modes $\ell\geq6$ in a spherical configuration requires an outer gap radius approximately twice the ablator outer radius (i.e., a case-to-capsule ratio $\rm{CCR}\equiv R_{\mathrm{converter}}/R_{\mathrm{capsule}}\simeq2$). This applies to the beam modes on OMEGA's LDD implosions ($10\leq\ell\leq30$). The CCR estimate is based on a view-factor method with straight-line radiation transport from the converter sphere to the concentric nested capsule. Our multidimensional \textsc{draco} simulations confirm the analytical results.

The OMEGA beam-port geometry suppresses all modes $\ell<10$ very effectively \cite{craxton2015direct}, assuming perfect beam alignment at target chamber center and ideal beam power balance. Higher $\ell$ modes (from the laser speckles) are expected to be rapidly smoothed even for $\rm{CCR}$s only slightly above unity. However, small $\rm{CCR}$s alongside target positioning can cause mode transfer from the converter to the capsule, affecting implosion performance. \textsc{Draco} simulations show that hydrocoupling can greatly reduce the fusion yield ($\rm{YOC}^{\rm{no}\mathrel{-}\upalpha}_{2D}\simeq 0.2$ to $0.35$). The resonances in the OMEGA laser's illumination spectrum provide the optimal radii where even the otherwise dominant $\ell=10$ is minimized in the initial nonuniformity (Fig.~\ref{fig:skupsky_modes}) \cite{skupsky1983uniformity,craxton2015direct,shvydky2023optimization}.

Therefore, the CCR can be readily optimized to permit efficient absorption of x rays while maximizing hydrodynamic efficiency. The standoff distance is now long enough that it is possible to hydrodynamically isolate the capsule from the motion of the converter. This long distance is combined with pulse shaping in such a way that the laser ablation of the capsule during the foot expels enough material to push the subcritical detritus from the converter, away from the capsule. Variable beam sizes (zooming) would provide greater flexibility in designing the converter shell, thus further improving the illumination uniformity. 

\textit{Numerical Simulations}--- Based on the above considerations, it is possible to design an $\alpha\sim2$ hybrid target for OMEGA that satisfies all requirements. The 1D metrics of the HSD design are shown in Table~\ref{tab:target_metrics} for two illumination schemes, with standard OMEGA fixed beam sizes (SG5-850) and with two-step zooming. The ``strong'' picket (Fig.~\ref{fig:pulse_shape}) requires slightly higher overall laser energy ($\sim$1.5~kJ) for the HSD design. A low-adiabat, bare LDD target version of the hybrid was also designed to get the same overall 1D performance and stability characteristics. Both targets are shown in Fig.~\ref{fig:LDD_HSD_targets}. The comparison to a standard bare target was done to highlight how the HSD target responds to imprint in comparison to typical LDD conditions.

The bare target uses the standard single-spot SG5-850 drive currently available on OMEGA for both the picket and main pulse. For the optimal HSD geometry with a 938~$\mu$m converter radius [point (2) of Fig.~\ref{fig:skupsky_modes}], corresponding to $\mathrm{CCR}\approx2$, these small beams overlap poorly on the converter and tile mid-mode structure onto its surface. The resulting x-ray hot spots produce substantial local variations in the initial spectrum. They also influence the structure of the converter detritus whose residual flows, although hydrodynamically decoupled from the capsule, create localized coronal nonuniformities that refract the main-drive light \cite{obenschain2002effects}. Reducing the initial converter illumination nonuniformity is therefore important.
\begin{table}
    \caption{\label{tab:target_metrics} 1D performance metrics. Paired SSD entries: On; Off.}
    \begin{ruledtabular}
        \begin{tabular}{cccccc}
            Design & Beam spot & SSD & Yield $(\times10^{14})$ & $\rho$R $(\rm{mg}/\rm{cm}^2)$ & $\alpha$\\[0.5ex]
            \hline\\[-2ex]
            Bare & SG5-850 & On & 8.7 & 215 & 2.0\\
            HSD & SG5-850 & On & 8.8 & 203 & 2.3\\
            HSD & ZPP & On; Off & 8.7 & 218 & 2.3\\
        \end{tabular}
    \end{ruledtabular}
\end{table}%

The second HSD design uses two-step zooming and places the converter more conservatively at $R_{\mathrm{conv}}=1070$~$\mu$m. The first step uses large spots, with $\eta\simeq3.9$ and $R_{95\%}\sim1200~\mu$m, to give converter uniformity comparable to point (1) of Fig.~\ref{fig:skupsky_modes} during the picket. The second step uses the standard SG5-850 spots, maintaining the capsule overlap used in standard OMEGA LDD targets during the foot. Zooming phase plates (ZPPs) \cite{craxton2015direct,kosc2015the,froula2013mitigation} would provide the required time-dependent spot change with a modest facility upgrade. This scheme does not improve the HSD energetics, since the post-picket main drive spot is identical for the bare and all HSD targets.

\begin{figure}[!htbp]
  \centering
  \begin{minipage}{\columnwidth}
    \centering
    \includegraphics[width=\textwidth]{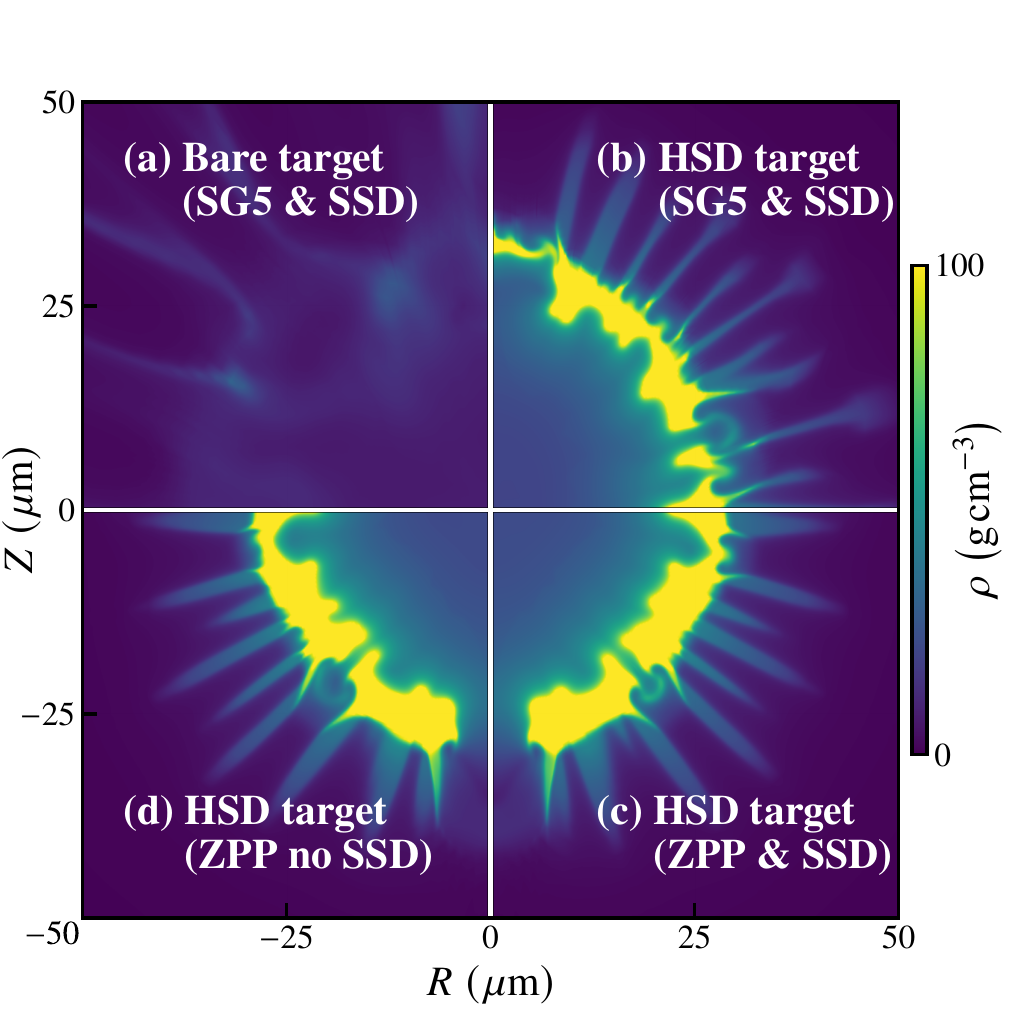}
    \caption{Bang-time plots of 2D \textsc{draco} simulations with beam-port geometry and imprint comparing (clockwise) insert (a) the $\alpha\simeq2$ LDD implosion and (b) its HSD equivalent, both with single spot illumination (SG5-850). The HSD design with ZPPs is shown with SSD on (c), and off (d). The HSD shows good shell integrity and little sensitivity to the SSD settings. Coronal refraction of early foot-seeded beam modes is the only source of perturbation in the HSD. Polar perforations are spot-dependent $r$--$z$ artifacts.} 
    \label{fig:2D_targets}
  \end{minipage}
\end{figure}

Two-dimensional \textsc{draco} simulations were carried out to evaluate the effects of beam-port geometry and imprint up to mode $\ell=100$ \cite{hu2023laser,cao2025impact}. The bare and HSD (with SG5-850's) targets are shown (a-b) near bang time in Fig.~\ref{fig:2D_targets}. The bare capsule on the top left is destroyed, whereas the hybrid on the top right shows good shell integrity. Perturbations in the hybrid shell arise primarily from beam modes ($\ell<50$) seeded at the beginning of the foot pulse (see Fig.~\ref{fig:pulse_shape}), well below the $\ell=100$ cutoff. The comparison of the HSD design using the SG5-850 distributed phase plates throughout the drive and ZPPs showed that refraction, is the main source of degradation. Laser refracts as it passes the perturbed converter detritus and imposes a perturbation onto the capsule (Fig.~\ref{fig:2D_targets}), but not imprint, which is successfully mitigated.
\begin{table}
    \caption{\label{tab:2D_metrics} 2D performance metrics with OMEGA's beam ports and imprint up to mode $\ell=100$. Paired SSD entries: On; Off.}
    \begin{ruledtabular}
        \begin{tabular}{ccccc}
            Design & Beam spot & SSD & Yield $(\times10^{14})$ & $\rho$R $(\rm{mg}/\rm{cm}^2)$\\[0.5ex]
            \hline\\[-2ex]
            Bare\footnote{Degradation: laser speckles (beam modes degrade yield by only $\sim$5\%).} & SG5-850 & On & 1.0 & 53\\
            HSD\footnote{Degradation: converter motion during the foot refracts the laser.} & SG5-850 & On & 5.5 & 176\\
            HSD$^{\mathrm{b}}$ & ZPP & On; Off & 6.7; 6.5 & 209; 207\\
        \end{tabular}
    \end{ruledtabular}
\end{table}

Optimizing the picket beam spots with ZPPs to reduce the initial illumination $\sigma_{_{\rm{RMS}}}$ on the converter weakens the residual mode amplitudes in the converter detritus. This increases the fusion yield from $5.5\times 10^{14}$ with fixed OMEGA phase plates to $6.7\times 10^{14}$ with ZPPs, a $\sim22\%$ improvement. The stagnated shell shape of the HSD target with ZPPs is shown in Fig.~\ref{fig:2D_targets}(c), where the simulations recover most of the shell integrity, accompanied by an increase in $\rho R$. Overall, the hybrid targets recover a substantial fraction of the 1D performance. The maximum normalized ``no-$\upalpha$'' Lawson parameter \cite{betti2015alpha,bose2016core}, $\chi^{\rm{no}\mathrel{-}\upalpha}_{2\rm{D}}\simeq 0.37$, is 85\% higher than the current OMEGA record \cite{gopalaswamy2024demonstration} ($\chi^{\rm{no}\mathrel{-}\upalpha}_{\Omega}\approx 0.2$). The 2D metrics in Table~\ref{tab:2D_metrics} show that the HSD targets outperform the bare LDD target in every respect. Since the Lawson parameter is proportional to $E_{Laser}^{1/3}$ \cite{bose2016core}, the projected 85\% enhancement lowers the laser energy required for ignition by at least a factor of four.

To quantify imprint mitigation, 2D HSD simulations with ZPPs were performed with and without laser SSD (smoothing by spectral dispersion) \cite{lle1999ssd}. As shown in Fig.~\ref{fig:2D_targets}(b,c) and Table~\ref{tab:2D_metrics}, both cases retain excellent stagnated-shell integrity, with no appreciable differences in structure or performance. This confirms that HSD strongly mitigates laser imprint and obviates the need for beam smoothing. These results indicate that HSD targets can substantially improve OMEGA implosions independently of SSD and even without zooming.

\textit{Conclusions}--- In summary, hybrid shock drive (HSD) target designs \cite{ceurvorst2020hybrid} provide a promising path to low-adiabat, high-convergence direct-drive implosions on OMEGA. The concept relies on a two-stage irradiation sequence: an external shell first converts the laser picket into a burst of x rays as it disassembles, and this x-ray flash launches a smooth initial shock into the cryogenic capsule nested inside. By setting the adiabat with radiation rather than direct laser illumination, the HSD scheme also forms a capsule corona before the main drive, shielding the capsule from laser imprint.

This Letter presents the first integrated simulations of a hybrid drive target, including both the x-ray source and the subsequent directly driven implosion. Hybrid targets for OMEGA with $\alpha\sim2$ were designed and compared with an equivalent laser-direct-drive target. The results show that HSD targets are robust to laser imprint and can perform well even without laser smoothing. Two-stage zooming gives the best overall uniformity by reducing both imprint and beam-geometry modes, but the central benefit does not depend on zooming: even the unzoomed HSD designs are expected to substantially outperform standard OMEGA targets at the same adiabat, and markedly reduce ignition-energy requirements, motivating high-gain hybrid LDD target designs relevant to IFE.

\textit{Acknowledgments}--- This material is based upon work supported by the Department of Energy (DOE) National Nuclear Security Administration under Award No. DE-NA0004144, the DOE Fusion Energy Science program under Award Numbers DE-SC0024456, DE-SC0024381, the STARFIRE collaboration, the University of Rochester, and the New York State Energy Research and Development Authority.

The support of DOE does not constitute an endorsement of the views expressed in this Letter. This report was prepared as an account of work sponsored by an agency of the US Government. Neither the US Government nor any agency thereof, nor any of their employees, makes any warranty, express or implied, or assumes any legal liability or responsibility for the accuracy, completeness, or usefulness of any information, apparatus, product, or process disclosed, or represents that its use would not infringe privately owned rights. Reference herein to any specific commercial product, process, or service by trade name, trademark, manufacturer, or otherwise does not necessarily constitute or imply its endorsement, recommendation, or favoring by the US Government or any agency thereof. The views and opinions of authors expressed herein do not necessarily state or reflect those of the US Government or any agency thereof.

\textit{Data availability}--- The data that support the findings of this study are available from the corresponding author upon reasonable request.

\bibliography{the_paper_bibliography}

\end{document}